# CVD Synthesis of Small-Diameter Single-Walled Carbon Nanotubes on Silicon

N. Arjmandi[1], P. Sasanpour[2], B. Rashidian[1,2*]

rashidia@sharif.edu

A simple process for chemical vapor deposition of ultra SD single wall carbon nanotubes has been developed. In this process, an iron nitrate nonahydrate solution in isopropyl alcohol with a concentration of $400\,\mu gr/mlit$ was used to catalyze nanoparticles formation on an oxidized silicon wafer. The oxide on the substrate was made of a thick layer of wet oxide sandwiched between tow thin layers of dry oxide. The process results in semiconducting single-walled carbon nanotubes (SWNTs) with diameter of less than 0.7nm and more than 1ev band gap energy, which are amongst the smallest diameters of SWNTs ever reported.

## 1. Introduction

Carbon nanotubes are currently the focus of intense research due to their unique properties and potential to impact broad areas of science and technology. The distinctive characteristics of carbon nanotubes arise from their size and atomic structure, for example a nanotube has high Young's modulus and tensile strength [1] and can be semi-metallic or semiconducting depending on its chirality and diameter [2]. Utilization of these properties in individual or ensembles of nanotubes has led to advanced scanning probes [3], transistors [4] and electron field-emission sources [5].

---

[1] Department of Electrical Engineering Sharif University of Technology
[2] Institute for NanoScience and NanoTechnology (INST) and Center of Excellence for Nanostructures Sharif University of Technology
[*] Corresponding Author

An important property of a semiconducting material, used for electronic devices, is its band gap energy, which, in turn, determines the electrical properties of the device. A semiconducting carbon nanotube's band gap energy is typically on the order of 0.5ev, which is low compared to over 1ev band gap of common semiconductors. Thus, due to inverse proportionality of semiconducting nanotubes' band gap to their diameter, synthesis of small diameter SWNTs is desired for many nanoelectronic applications such as carbon nanotube field effect transistors (CNTFETs). In addition, development of controlled synthesis methods to obtain nanotubes with predetermined diameter, chirality, position, and alignment is an important and viable route to potential applications of CNTs. In addition, the most applicable method to control the diameter and specially position and alignment of nanotubes is chemical vapor deposition (CVD). We investigated CVD synthesis of carbon nanotubes and designed and developed the necessary equipment. In the next step we modified and adjusted an established process to perform CVD synthesis of ultra small diameter nanotubes.

## 2. Equipment

The main equipment we used is a horizontal gas phase reactor like many other systems for CVD synthesis of carbon nanotubes. The reaction chamber is a quartz tube, with a length of 130 cm and diameter of 6 cm (figure 1). During the growth process, we need to heat the tube up to 750ºc and this temperature must be stable and controllable over a distance of few decimeters in the middle of the tube. In addition, its inside should be isolated from the atmosphere carefully, because any oxygen leak can cause an explosion in the tube. Reactants are feed from the gas

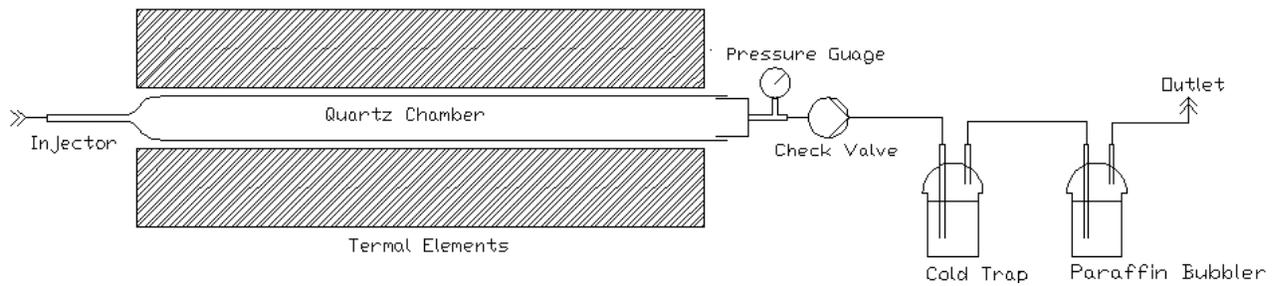

Figure1. Schematic diagram of reaction chamber.

cabinets through a gas circuit (figure 2) that controls the flow of each species and provides safety arrangements for using the highly explosive gases. For CVD synthesis of carbon nanotubes, usually explosive gases are used at a temperature well above their autoignition temperature and safety regulations need to be considered carefully. The most important parts for this are; flash back arrestors that prevent the flash back of gases to the cylinders and prevent the spreading of explosion to them. Also a water bubbler and a paraffin trap are used at the exhaust to cool and neutralize the exhaust gases.

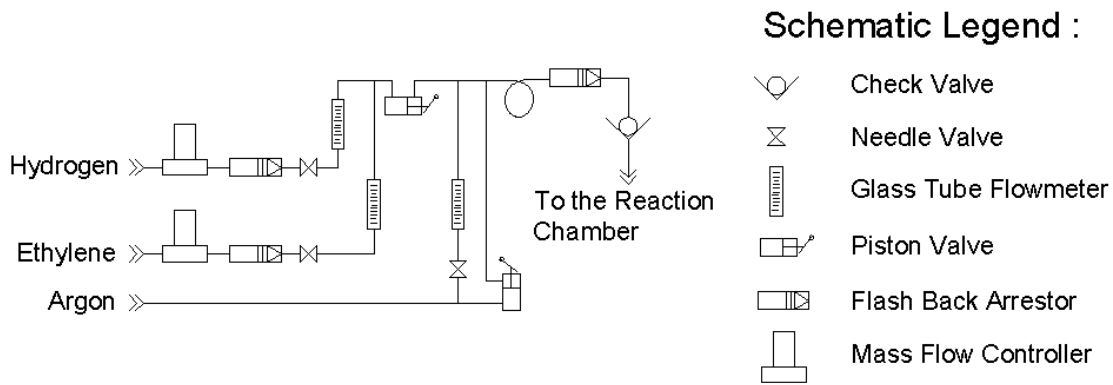

Figure2. Reactor's schematic diagram of gas circuit.

## 3. Process

We slightly modified an established easy process for CVD synthesis of SWNTs [6] to produce SWNTs with the desired diameter. In this process p-type, <111>, polished silicon wafers were cut to 4mm×8mm samples and cleaned using a modified version of standard semiconductor wet cleaning, which was done in the following steps: ultrasonication and then washing by TCE, acetone, methanol, DI Water, piranha (H2SO4:H2O2 = 4:1, v/v), DI Water, dilute HF solution (%5), DI Water, SC1 (NH4OH:H2O:H2O2 = 1:5:1, v/v), SC2 (HCl:H2O:H2O2 = 1:6:1, v/v), dilute HF solution (%5), DI Water and drying under N2 Spray. We sandwiched the wet oxide between two layers of thin dry oxides grown on the substrate to have flat and high quality surface in a short time. For this manner we used 15min dry oxidation at 1100ºC, followed by about 30min wet and 15min dry oxidation. Thus the total thickness of the oxide was determined to be about 500nm. A solution of iron nitrate nonahydrate in isopropyl alcohol ($400\,\mu gr/mlit$) was prepared. Si substrates were dipped in the solution for 10sec, rinsed 10 times in n-hexane and dried in air. For CVD, samples were placed in the reactor and heated to 750 ºC under argon atmosphere with a flow of 600sccm. The samples were then annealed for 15min at 750ºC under the flow of argon (600sccm) and hydrogen (400sccm). CNT growth was carried out by adding 10sccm Ethylene for 5min. Then hydrogen and ethylene's flow were turned off, and the reaction chamber was cooled down over night under argon flow and the samples were brought out when the chamber's temperature was below 200ºC.

## 3. Results and Discussion

Some processes based on ferric nitrate nanoparticle formation on silicon dioxide have previously been established to synthesize relatively large diameter (around 5nm) SWNTs [7], and this study demonstrates the possibility of synthesizing ultra small diameter SWNTs using ferric nitrate catalyst nanoparticles. To achieve this goal, we used a much lower Fe(NO3)3.9H2O/isopropyl alcohol concentration and a thick wet oxide sandwiched between tow layers of thin dry oxide on the surface of silicon wafer instead of only wet oxide. In addition, it is essential in this process to cool down as grown SWNTs in an inert atmosphere to temperatures lower than 200ºC. Because ultra small diameter SWNTs burn with oxygen at higher temperatures (the exact burning temperature of a SWNT strongly depends on its defect density and diameter and it is usually between 200ºC and 400ºC [8]).

Prepared samples were characterized using atomic force microscopy (AFM). For this purpose, we started with a large area scan in AC mode and continued with low- and high-resolution scans in DC mode (figures 3 and 4).

Using high-resolution scans, we can extract the precise values of carbon nanotube diameter and calculate their chirality vector. Figure 4.b shows various cross sectional profiles extracted from a typical high resolution scan of the nanotubes. As this figure demonstrates, their diameter is less than 0.7nm. Inasmuch as it is proved that the SWNTs' diameter cannot be smaller than about 0.4nm in room temperature (It's proved theoretically that it's not possible to synthesize SWNTs slightly with diameter smaller than 0.4nm [9,10] while one can find 0.3nm thick nanotubes as the most inner layer of MWNTs (multi wall carbon nanotubes) [11], thus these are one of the thinnest possible single wall carbon nanotubes, while being made using iron nitrate solution for nanoparticle formation, which is an inexpensive and useful method for nanoelectronic applications.

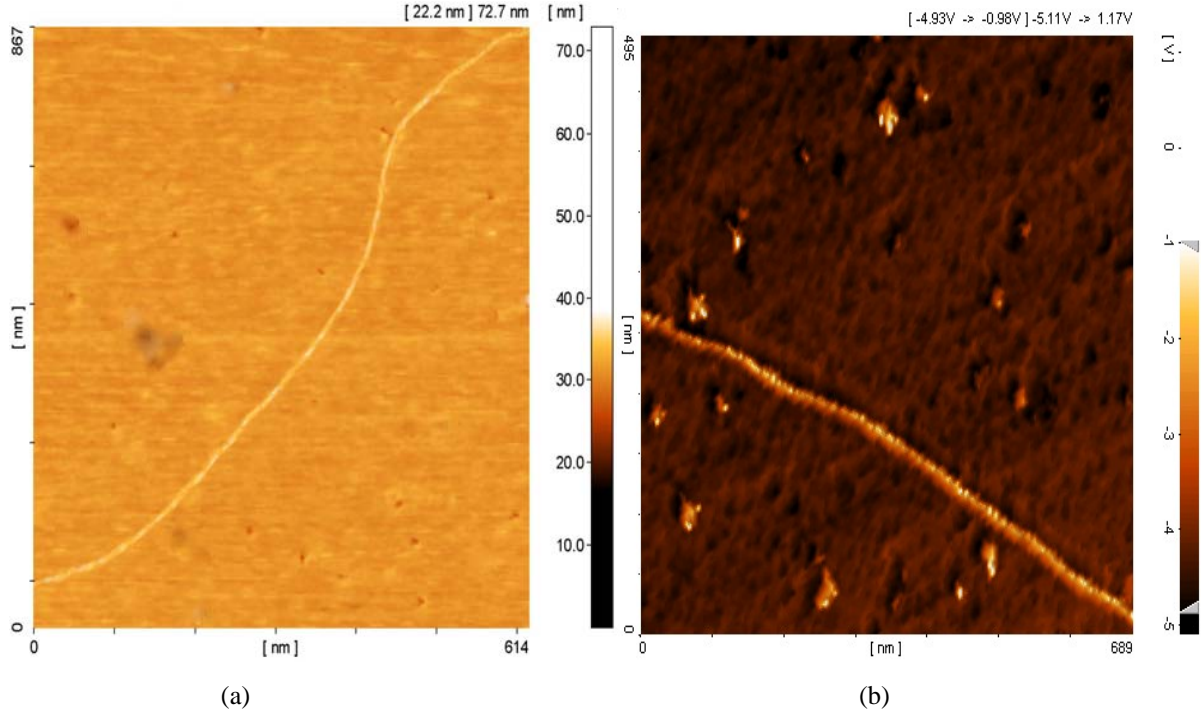

(a)          (b)

Figure 3. A high resolution scan of the prepared sample a) A single wall carbon nanotubes in a topographic image. b) The phase image of a SWNT shows the difference between SWNT and substrate's type of atoms.

The chirality vector of the synthesized nanotubes was calculated from high-resolution scans using the established methods. Figure 5 shows a typical synthesized nanotube, the observed chirality angle of the SWNT is about 43º (the angle between the parallel lines on the SWNT and SWNT's axle). By applying the correction factors to the image, the accurate angle was calculated to be about 8º (the correction factor is $\frac{1}{2}\sqrt{\frac{D}{r+\Delta}}$, where D is the nanotubes diameter, r is the AFM tip's radius and $\Delta$ is the distance between the tip and the nanotube [12]). Using the extracted value of the nanotube diameter (0.69nm) one can calculate the chirality vector that is (9,2). Thus, it is a semiconducting nanotube with a band gap of over 1ev ($E_g \approx \frac{0.705nm}{D}$, where $E_g$ is the band gap and $D$ is the nanotube's diameter [13]).

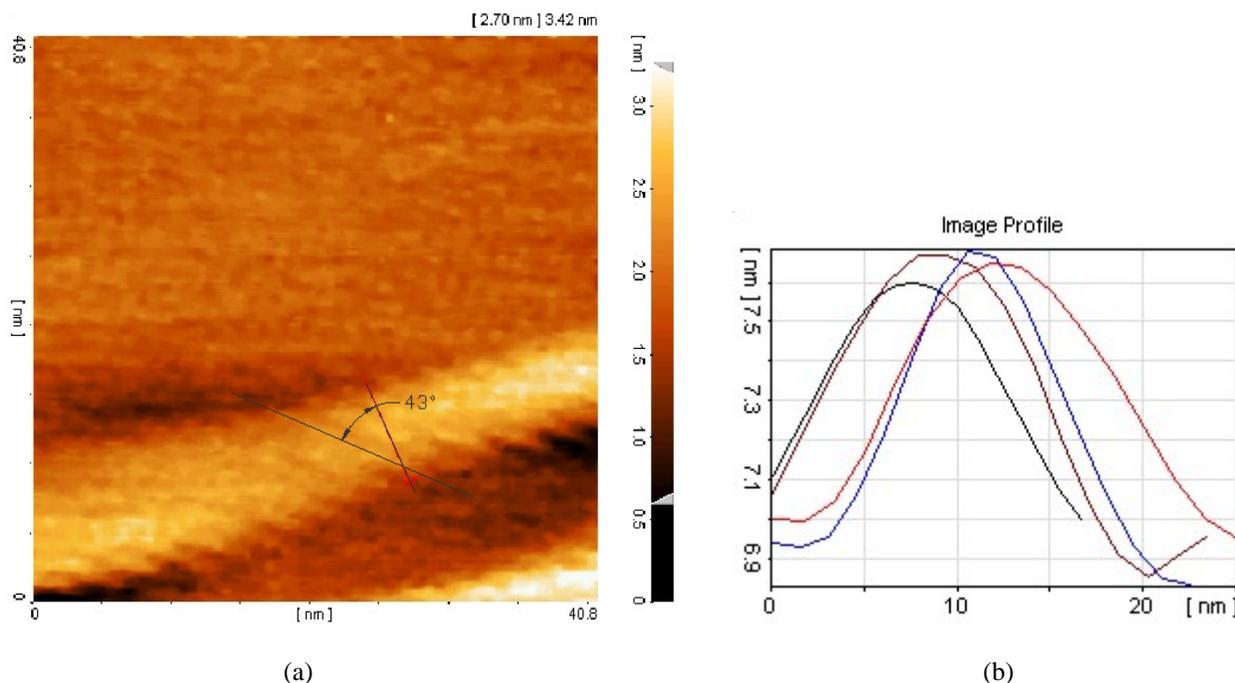

(a)            (b)

Figure 5. a) A very high resolution scan of a single wall carbon nanotube. Chirality and diameter can extract from this picture. b) Cross sectional profile of the synthesized nanotubes. The nanotubes diameter can extract from he height of these curves that is about 0.69nm. In addition if we deconvolve the cross sectional curves with the tip's profile we will fine the diameter to be about 0.68nm. Thus we are sure that the SWNTs haven't flattened (maybe due to its small diameter it's more rigid) and their diameter is less than 0.7nm.

## 5. Conclusion

A simple method was introduced for CVD synthesis of ultra small diameter semiconducting single-walled carbon nanotubes (SWNTs). In this method, we placed iron nitrate catalyst nanoparticles on oxidized silicon substrate using a dilute solution of iron nitrate nonahydrate in isopropyl alcohol and a thick wet oxide sandwiched between tow layers of thin dry oxides. The process was carried out like common carbon nanotube growth processes by using ethylene as the carbon containing reactant, hydrogen as the catalyst depoisoner, and paying special attention to the cool-down procedure to avoid burning the ultra small diameter nanotubes. It worth to mention that if we define the yield of this process as the ratio of number of SWNTs to the number of nanoparticles it will be quite low and at area which was covered by hundreds of nanoparticles a few SWNTs observed.

Atomic force microscopy (AFM) measurements showed that the process resulted in ultra small diameter semiconducting single-walled carbon nanotubes with diameters of less than 0.7 nm, produced by iron nitrate for the first time. These diameters would correspond to band gap energies of more than 1ev (as it calculated from the FM images).

We think the very low concentration of the Iron nitrate together with the appropriate surface preparation were led to very small catalysts nanoparticles which in combination with appropriate thermal conditions and gas flow rates; resulted in ultra small diameter SWNTs. As we found it's also very important to cool down the chamber under an inert ambient to the 200°C, maybe because the very small diameter SWNTs burn so easily in higher temperatures.